\documentclass{ws-procs9x6}
\usepackage{latexsym}
\usepackage{epsfig,graphics}

\begin{document}

\title{
Baryon vector and axial-vector currents in the $1/N_c$ expansion
}

\author{
Ra\'ul Acosta and Rub\'en Flores-Mendieta
}

\address{
Instituto de F{\'\i}sica, Universidad Aut\'onoma de San Luis Potos{\'\i}, \\
\'Alvaro Obreg\'on 64, Zona Centro, \\
San Luis Potos{\'\i}, S.L.P.\ 78000, Mexico
}

\maketitle


\abstracts{
The derivation of the $1/N_c$ expansions for the baryon vector and axial-vector currents, in the SU(3) symmetry limit and
including perturbative SU(3) breaking, is reviewed. Symmetry breaking effects in the hyperon semileptonic decay form factors
are reanalyzed. Fits to experimental data yield corrections to the leading vector and axial-vector form factors consistent
with expectations. The values of the Cabibbo-Kobayashi-Maskawa matrix elements $V_{ud}$ and $V_{us}$ also extracted from the
analysis are comparable to the ones recommended by the Particle Data Group.
}

\section{Introduction}

A powerful method which has been decisive in the understanding of low-energy QCD hadron dynamics is the $1/N_c$
expansion.\cite{th,witt} This method promotes QCD to an SU($N_c$) non-Abelian gauge theory, where $N_c$ is the number of
colors. In the limit $N_c \to \infty$ it has been shown that the baryon sector possesses a contracted SU$(2F)$ spin-favor
symmetry, where $F$ is the number of light quark flavors.\cite{dm,gs} Large-$N_c$ baryons form irreducible representations of
the spin-flavor algebra\cite{djm} and their static properties can be computed in a systematic expansion in $1/N_c$.
Outstanding evidence of the predictions of the $1/N_c$ expansion can be found in the analysis of baryon masses,\cite{djm,jl}
magnetic moments,\cite{djm,dai,lmrw} and axial-vector and vector currents,\cite{djm,dai,rfm98} which are in good agreement
with the experimental data.

An interesting topic that can be tackled in the context of the $1/N_c$ expansion is the evaluation of flavor SU(3) symmetry
breaking (SB) corrections in the hyperon $\beta$-decay form factors. A deep understanding of these corrections is indeed
important for a precise determination of the Cabibbo-Kobayashi-Maskawa (CKM) matrix elements $V_{ud}$ and $V_{us}$ from
hyperon semileptonic decays (HSD). Currently, $V_{ud}$ can be precisely obtained from superallowed $0^+ \to 0^+$ $\beta$
decays whereas $V_{us}$ is presumably more reliably extracted from $K_{e3}$ rather than HSD\cite{part} because of the larger
theoretical uncertainties due to first-order SU(3) SB in the axial-vector form factors.

Here we come back to the study of HSD form factors in a combined expansion in $1/N_c$ and SU(3) flavor SB by following the
lines of previous works,\cite{djm,dai,rfm98,rfm04} with the introduction of some variants. First (Sec.~2) we review the
construction of the baryon vector and axial-vector currents whose matrix elements between SU(6) symmetric states yield the
actual values of the vector and axial-vector form factors at $q^2=0$, namely $f_i$ and $g_i$, with $i=1,2$. We also briefly
review (Sec.~3) the existent information in HSD. We then proceed (Sec.~4) to perform a comparison of the theoretical
expressions with the available experimental data for the decay rates, angular correlations and angular spin-asymmetry
coefficients of the octet baryons, and for the widths (converted to axial-vector couplings by using the Goldberger-Treiman
relation) of the decuplet baryons. Finally we discuss our findings in a closing section (Sec.~5).

\section{Operator analysis}

The theoretical groundwork on the spin-flavor structure of large-$N_c$ baryons has been established,\cite{djm} so we will
only provide a brief summary of the main results here. For the analysis of HSD data, however, we follow the lines of previous
works.\cite{dai,rfm98,rfm04}

The $1/N_c$ expansion using quark operators as the operator basis\cite{cgo,lm,djm} provides a framework for studying the
spin-flavor structure of baryons. For $F=3$, the lowest lying baryon states fall into a representation of the spin-flavor
group SU(6). When $N_c=3$, this corresponds to the familiar {\bf 56} dimensional representation of SU(6).

A complete set of operators can be constructed using the zero-body operator $\mathbb I$ and the one-body operators\cite{djm}
\begin{eqnarray}
J^i & = & q^\dagger \left(\frac{\sigma^i}{2} \otimes {\mathbb I} \right) q \qquad (1, 1), \nonumber \\
T^a & = & q^\dagger \left( {\mathbb I} \otimes \frac{\lambda^a}{2} \right) q \qquad (0, 8), \label{eq:gen} \\
G^{ia} & = & q^\dagger \left( \frac{\sigma^i}{2} \otimes \frac{\lambda^a}{2} \right) q \qquad (1, 8), \nonumber
\end{eqnarray}
where $J^i$ are the baryon spin generators, $T^a$ are the baryon flavor generators, and $G^{ia}$ are the baryon spin-flavor
generators. The transformation properties of these operators under SU(2)$\times$SU(3) are indicated as $(j,d)$ in
Eq.~(\ref{eq:gen}), where $j$ is the spin and $d$ is the dimension of the SU(3) flavor representation.

Any QCD one-body operator transforming according to a given SU(2)$\times$SU(3) representation has a $1/N_c$ expansion of the
form\cite{djm}
\begin{eqnarray}
{\mathcal O}_{\rm QCD} = \sum_{n=0}^{N_c} c_n \frac{1}{N_c^{n-1}}{\mathcal O}_n \label{eq:genex},
\end{eqnarray}
where $c_n$ are unknown coefficients which have power series expansions in $1/N_c$ beginning at order unity. The sum in
Eq.~(\ref{eq:genex}) extends over all possible independent $n$-body operators ${\mathcal O}_n$ with the same spin and flavor
quantum numbers as ${\mathcal O}_{\rm QCD}$. By using operator identities\cite{djm} it is always possible to reduce the
operator basis to a set of independent operators.

We now proceed to derive the $1/N_c$ expansions for the HSD amplitudes to first order in flavor SB, and to leading order in
$1/N_c$ for most of the form factors, except for $f_1$ which is protected by the Ademollo-Gatto theorem against SU(3)
breaking corrections to lowest order in $(m_s-\hat{m})$; therefore we need to include second-order flavor SB corrections in
$f_1$. The contributions of $f_2$ and $g_2$ to the decay amplitudes are suppressed by the momentum transfer. In the symmetry
limit the hyperon masses are degenerate and then such contributions vanish. Thus, the first-order SB to $f_2$ and $g_2$
actually contribute to second order in the decay amplitude and can be neglected. The axial form factor $g_1$ is computed to
first order in SB. Finally, the form factors $f_3$ and $g_3$, for electron or positron emission, have contributions
proportional to the electron mass squared so we can safely neglect them.

\subsection{Vector form factor $f_1$}

To begin with, let us write down the $1/N_c$ expansion for the baryon vector current in the SU(3) flavor symmetry limit. At
$q^2 = 0$, the hyperon matrix elements for the vector current are given by the matrix elements of the associated SU(3)
generator. Let $V^{0a}$ denote the flavor octet baryon charge\cite{rfm98}
\begin{eqnarray}
V^{0a} = \left< B^\prime \left|{\left(\overline {q} \gamma^0 \frac{\lambda^a}{2}q\right)}_{\rm QCD} \right| B \right>,
\end{eqnarray}
whose matrix elements between SU(6) symmetric states yield $f_1$. In the above expression the subscript QCD emphasizes that
$\overline {q}$ and $q$ are QCD quark fields, not the quark creation and annihilation operators of the quark representation.
$V^{0a}$ is spin-0 and a flavor octet, so it transforms as (0,8) under SU(2)$\times$SU(3).

The $1/N_c$ expansion for a (0,8) operator has been already obtained.\cite{jl} By using operator reduction rules it is found
that only $n$-body operators with a single factor of either $T^a$ or $G^{ia}$ appear. The allowed one- and two-body operators
are given by
\begin{eqnarray}
{\mathcal O}_1^a = T^a, \qquad \qquad {\mathcal O}_2^a = \{J^i,G^{ia}\}, \nonumber
\end{eqnarray}
and the remaining operators $(n\geq 3$) are obtained as ${\mathcal O}_{n+2} = \{J^2, {\mathcal O}_n\}$. The $1/N_c$ expansion
of $V^{0a}$ is then
\begin{eqnarray}
V^{0a} = \sum_{n=1}^{N_c} c_n \frac{1}{N_c^{n-1}}{\mathcal O}_n^a \label{eq:vcoup}.
\end{eqnarray}
At $q^2=0$ $V^{0a}$ is the generator of SU(3) symmetry transformations so
\begin{eqnarray}
c_1 = 1, \qquad c_n = 0,\ n > 1. \nonumber
\end{eqnarray}
Therefore, in the limit of exact SU(3) flavor symmetry one has
\begin{eqnarray}
V^{0a} = T^a, \label{eq:vafin}
\end{eqnarray}
to all orders in the $1/N_c$ expansion.

\subsection{Vector form factor with perturbative SU(3) breaking}

Flavor SU(3) SB is due to the strange quark mass in QCD and transforms as a flavor octet. Constructing the most general
$1/N_c$ expansion for $V^{0a}$ up to second-order in symmetry breaking requires to consider all the possible spin-0
SU(2)$\times$SU(3) representations of the quark operators contained in the SU(6) representations {\bf 1}, {\bf 35}, {\bf 405}
and {\bf 2695} allowed by time reversal invariance, {\it i.e.}\ $(0,1)$, $(0,8)$, $(0,27)$, $(0,64)$, and
$(0,10+\overline {10})$, since the baryon $1/N_c$ expansion extends only to three-body operators if we restrict ourselves to
physical baryon states. This problem has been already solved\cite{jl} and the results can be summarized as follows.

For a $(0,1)$ operator the $1/N_c$ expansion starts with the zero-body operator ${\mathcal O}_0 = {\mathbb I}$ and the
remaining operators are obtained as ${\mathcal O}_{2m} = \{J^2, {\mathcal O}_{2m-2}\}$, for $m\geq 1$.

The $1/N_c$ expansion for a $(0,8)$ operator has the same form as Eq.~(\ref{eq:vcoup}) whereas for a $(0,27)$ operator the
$1/N_c$ expansion contains the two- and three-body operators
\begin{eqnarray}
{\mathcal O}_2^{ab} = \{T^a, T^b\}, \quad \qquad  {\mathcal O}_3^{ab} = \{T^a, \{J^i, G^{ib}\}\} + \{T^b, \{J^i, G^{ia}\}\},
\nonumber
\end{eqnarray}
where the flavor singlet and octet components of the above operators are subtracted off. For a $(0,64)$ operator, the $1/N_c$
expansion starts with a single three-body operator
\begin{eqnarray}
{\mathcal O}_3^{abc} = \{T^a, \{T^b, T^c \}\}, \nonumber
\end{eqnarray}
where the singlet, octet and 27 components are subtracted off to leave only the 64 component. Finally, for a
$(0,10+\overline{10})$ operator one obtains
\begin{eqnarray}
{\mathcal O}_3^{ab} = \{T^a, \{J^i,G^{ib}\}\} - \{T^b,\{J^i,G^{ia}\}\}. \nonumber
\end{eqnarray}

First-order SB terms in $V^{0a}$ are given by setting one free flavor index equal to 8 in the appropriate operators. At
second-order in SB, two free flavor indices are set equal to 8. One thus obtains\cite{rfm98}
\begin{eqnarray}
V^{0a} + \delta V^{0a} & = & (1 + \epsilon a_1) T^a + \epsilon a_2 \frac{1}{N_c}\{J^i,G^{ia}\} + \epsilon a_3\frac{1}{N_c^2}
\{J^2,T^a\} \nonumber \\
&   & \mbox{} + \epsilon b_1 d^{ab8} T^b + \epsilon b_2 \frac{1}{N_c} d^{ab8}\{J^i,G^{ib}\} \nonumber \\
&   & \mbox{} + \epsilon b_3 \frac{1}{N_c^2} d^{ab8}\{J^2,T^b\} + \epsilon a_4\frac{1}{N_c} \{T^a,T^8\} \nonumber \\
&   & \mbox{} + \epsilon a_5 \frac{1}{N_c^2} \left( \{T^a, \{J^i, G^{i8}\}\} + \{T^8, \{J^i,G^{ia}\}\} \right) \nonumber \\
&   & \mbox{} + \epsilon a_6 \frac{1}{N_c^2} \left( \{T^a, \{J^i, G^{i8}\}\} - \{T^8, \{J^i,G^{ia}\}\} \right) \nonumber \\
&   & \mbox{} + \epsilon^2 b_4 \frac{1}{N_c} d^{ab8}\{T^b,T^8\} + \epsilon^2 a_7 \frac{1}{N_c^2} \{T^a,\{T^8,T^8\}\}
\nonumber \\
&   & \mbox{} + \epsilon^2 b_5 \frac{1}{N_c^2} d^{ab8} \left( \{T^b, \{J^i, G^{i8}\}\} + \{T^8, \{J^i,G^{ib}\}\} \right)
\nonumber \\
&   & \mbox{} + \epsilon^2 b_6 \frac{1}{N_c^2} d^{ab8} \left( \{T^b, \{J^i, G^{i8}\}\} - \{T^8, \{J^i,G^{ib}\}\} \right).
\label{eq:vazero}
\end{eqnarray}
Here $\epsilon$ is a measure of SU(3) breaking. Equation~(\ref{eq:vazero}) has been written in such a way that there is no SB
for the $\Delta S=0$ weak decays, since isospin symmetry is not broken by the strange quark mass.

If we introduce the number of strange quarks, $N_s$, and the strange quark spin, $J_s^i$, which are defined through\cite{djm}
\begin{eqnarray}
T^8 = \frac{1}{2\sqrt 3} (N_c - 3 N_s), \qquad \qquad G^{i8} = \frac{1}{2\sqrt 3}(J^i - 3J_s^i), \nonumber
\end{eqnarray}
then Eq.~(\ref{eq:vazero}) can be given, after rearranging terms and absorbing factors of $N_c^{-1}$ and $N_c^{-2}$, as
\begin{equation}
V^{0a} = T^a, \label{eq:vds0}
\end{equation}
for $\Delta S=0$ HSD, and
\begin{eqnarray}
V^{0a} = (1 + v_1)T^a + v_2 \{T^a, N_s\} + v_3 \{T^a,-I^2+J_s^2\}, \label{eq:vds1}
\end{eqnarray}
for $|\Delta S| = 1$ HSD. Here $I$ is the isospin. The baryons are eigenstates of $J^2$, $I^2$, $J_s^2$, and $N_s$, so the
matrix elements of Eq.~(\ref{eq:vds1}) can be computed rather easily.\cite{rfm98} Thus, for any process, the matrix elements
of $V^{0a}$ are given as the sum of the matrix elements of the operators involved in the expansion times their respective
coefficient.

\subsection{Axial-vector form factor $g_1$}

The $1/N_c$ expansion for the axial-vector current $A^{ia}$, whose matrix elements yield $g_1$, has been already
obtained.\cite{djm} It reads
\begin{eqnarray}
\frac12 A^{ia} & = & a G^{ia} + bJ^i T^a + \Delta^a(c_1 G^{ia} + c_2 J^iT^a) + c_3 \{G^{ia},N_s\} + c_4\{T^a, J_s^i\}
\nonumber \\
&   & \mbox{} + \frac{1}{\sqrt 3} \delta^{a8}W^i - \frac{d}{2} ( \{J^2, G^{ia}\} - \frac12 \{J^i,\{J^j,G^{ja}\}\} ),
\label{eq:vafit}
\end{eqnarray}
where
\begin{eqnarray}
W^i & = & (c_4 - 2c_1)J_s^i + (c_3 - 2c_2)N_s J^i - 3(c_3 + c_4)N_s J_s^i, \nonumber
\end{eqnarray}
and $\Delta^a = 1$ for $a = 4,5,6$, or 7 and vanishes otherwise. In order to avoid the mixing between SB effects and $1/N_c$
corrections in the symmetric couplings a term proportional to $d$ is added.\cite{dai} This will also allow the SU(3)
symmetric parameters $D$, $F$, and ${\mathcal C}$ to have arbitrary values.

For definiteness, in the present analysis Eq.~(\ref{eq:vafit}) can be split into two contributions, according to the physical
processes we are concerned with. Thus one has,
\begin{equation}
\frac12 A^{ia} = a G^{ia} + bJ^i T^a - \frac{d}{2} ( \{J^2, G^{ia}\} - \frac12 \{J^i,\{J^j,G^{ja}\}\} ), \label{eq:ds0f}
\end{equation}
for $\Delta S=0$ processes,
and
\begin{equation}
\frac12 A^{ia} = a^\prime G^{ia} + b^\prime J^i T^a + c_3 \{G^{ia},N_s\} + c_4\{T^a, J_s^i\}, \label{eq:ds1f}
\end{equation}
for $|\Delta S|=1$ processes.

In the first case the couplings have been parametrized in such a way that only the parameters $a$, $b$, and $d$ contribute to
strangeness-zero processes, namely the $\Delta S=0$ HSD and the strong decays $\Delta \to N \pi$, $\Xi^*\to\Lambda \pi$,
$\Sigma^*\to \Sigma \pi$, and $\Xi^* \to \Xi \pi$. In the second case we ignore the contribution of the term proportional to
$d$, which results in redefinitions of the parameters $a$ and $b$ of Eq.~(\ref{eq:vafit}) into $a^\prime$ and $b^\prime$,
which absorb the terms $c_1$ and $c_2$, respectively, of the original expansion. Equations (\ref{eq:ds0f}) and
(\ref{eq:ds1f}) are the ones used in the fits to the experimental data.

\subsection{Weak magnetism form factor $f_2$}

In the limit of exact SU(3) flavor symmetry the form factor $f_2$ is determined by two invariants, $m_1$ and $m_2$, which can
be extracted from the anomalous magnetic moments of the nucleons.\cite{dai} The magnetic moment is a spin-1 octet operator so
it has a $1/N_c$ expansion identical in structure to $A^{ia}$. For convenience $m_1$ and $m_2$ are defined
through\cite{djm,dai}
\begin{equation}
M^i = m_1 G^{iQ} + m_2 J^i T^Q,
\end{equation}
where $Q$ represents the SU(3) generator (the electric charge), so $G^{iQ} \equiv G^{i3}+G^{i8}/\sqrt 3$, and $T^Q \equiv T^3
+ T^8/\sqrt 3$.

Previous works (see e.g.\ Ref.~[\refcite{rfm98}] and references therein) have shown that reasonable shifts from the SU(3)
predictions of $f_2$ have no perceptible effects upon $\chi^2$ or $g_1$ in a global fit to experimental data. We therefore
follow these works and determine $f_2$ with the best fit values\cite{dai} $m_1=2.87$ and $m_2=-0.77$.

\subsection{Weak electricity form factor $g_2$}

In the SU(3) flavor symmetry limit, the form factor $g_2$ vanishes, so that it is proportional to SU(3) SB at leading order.
$g_2$ transforms oppositely to $g_1$ and $f_2$ under time-reversal, and therefore has a different $1/N_c$ operator expansion.

Let $W^{ia}$ be the operator whose matrix elements yield $g_2$. At first order in SU(3) SB, the contribution to $g_2$
transforms as $(1,8)$ and $(1,10-\overline {10})$ under spin and flavor. The $(1,8)$ expansion reads\cite{rfm98}
\begin{eqnarray}
\delta W_8^{ia} \propto i b_1 f^{ab8} G^{ib} + i b_2 f^{ab8}\frac{J^iT^b}{N_c}, \label{eq:dw}
\end{eqnarray}
whereas the $i(10-\overline {10})$ expansion is
\begin{eqnarray}
\delta W_{10-\overline{10}}^{ia} \propto i f^{8cg}d^{ach} \left( \{ G^{ig},T^h \} - \{ T^g, G^{ih} \} \right). \label{eq:dwb}
\end{eqnarray}
For any process, the matrix elements of $W^{ia}$ can be given as a sum of the three parameters $b_{1-3}$ times the operator
matrix elements involved in Eqs.~(\ref{eq:dw}) and (\ref{eq:dwb}). Previous analyses,\cite{rfm98} however, have concluded
that the HSD data are not accurate enough for an extraction of the small $g_2$-dependence of the decay amplitudes. In
consequence, we also take the value $g_2=0$ in this work in order to have a consistent analysis.

\section{HSD data}

\subsection{Integrated observables}

The total decay rate $R$ and angular correlation and asymmetry coefficients are some integrated observables in HSD which can
be used when experiments have low statistics and no analysis of the differential decay rate is possible. These observables
are defined using only kinematics and no particular theoretical approach is assumed in their definitions. The charged
lepton-neutrino angular correlation coefficient, for instance, is defined as
\begin{equation}
\alpha_{\ell\nu} = 2 \frac{N(\Theta_{\ell\nu} < \pi/2) - N(\Theta_{\ell\nu} > \pi/2)}{N(\Theta_{\ell\nu} < \pi/2)
+ N(\Theta_{\ell\nu}> \pi/2)},
\end{equation}
where $N(\Theta_{\ell\nu} < \pi/2)$ $[N(\Theta_{\ell\nu} > \pi/2)]$ is the number of charged lepton-neutrino pairs emitted in
directions that make an angle between them smaller [greater] than $\pi/2$. Similar expressions are obtained for the charged
lepton $\alpha_\ell$, neutrino $\alpha_\nu$, and emitted hyperon $\alpha_B$ asymmetry coefficients, where $\Theta_\ell$,
$\Theta_\nu$, and $\Theta_B$ are this time the angles between the $\ell$, $\nu$, and $B_2$ directions and the polarization of
$B_1$, respectively. If the polarization of the emitted hyperon is observed, two more asymmetry coefficients, $A$ and $B$,
can be defined.

The unpolarized total decay rate $R^0$ is a quadratic function of the form factors and can be written in the most general
form as\footnote{A superscript 0 on a given observable indicates that no radiative corrections have been incorporated into
it.}
\begin{equation}
R^0 = \sum_{i\leq j=1}^6 a_{ij}^R f_i f_j + \sum_{i\leq j=1}^6 b_{ij}^R (f_i \lambda_{f_j} + f_j \lambda_{f_i}).
\label{eq:rnum}
\end{equation}
Here, for the sake of shortening Eq.~(\ref{eq:rnum}), we have momentarily redefined $g_1=f_4$, $g_2 =f_5$, $g_3=f_6$,
$\lambda_{g_1}=\lambda_{f_4}$, $\lambda_{g_2}=\lambda_{f_5}$, and $\lambda_{g_3}=\lambda_{f_6}$. Each sum in
Eq.~(\ref{eq:rnum}) contains 21 terms due to the restriction $i\leq j$. Although the form factors have been assumed to be
constant, their $q^2$-dependence cannot always be neglected because they may contribute to the observables. We have already
pointed out that the $q^2$-dependence of $f_2$ and $g_2$ can be ignored because they already contribute to order
${\mathcal O}(q)$ to the decay rate. For $f_1(q^2)$ and $g_1(q^2)$, however, non-negligible contributions can be obtained
with a linear expansion in $q^2$, namely,
\begin{equation}
f_1(q^2) = f_1(0) + \frac{q^2}{M_1^2} \lambda_1^f, \qquad \qquad g_1(q^2) = g_1(0) + \frac{q^2}{M_1^2} \lambda_1^g,
\nonumber
\end{equation}
where the slope parameters $\lambda_1^f$ and $\lambda_1^g$ are both of order unity.\cite{gk}

Similar expressions to Eq.~(\ref{eq:rnum}) also hold for the products $R^0 \alpha^0$, where $\alpha^0$ is any of the angular
coefficients defined above. Once $R^0$ and $R^0 \alpha^0$ are determined, $\alpha^0$ is obtained straightforwardly. Updated
values of the coefficients $a_{ij}^R$ and $a_{ij}^{R\alpha}$ can be found elsewhere.\cite{rfm04,acosta}

\subsection{Experimental data in HSD}

The measured quantities in HSD are the total decay rate $R$, angular correlation coefficients $\alpha_{e\nu}$, and angular
spin-asymmetry coefficients $\alpha_e$, $\alpha_\nu$, $\alpha_B$, $A$, and $B$. Often the measured $g_1/f_1$ ratios are also
presented. This latter set, however, is not as rich as the former and will not be used in the current analysis unless noted
otherwise. These experimental data\cite{part} for HSD are listed in Tables \ref{t:tab1} and \ref{t:tab2}. Besides, the
experimentally measured quantity for the decuplet baryons is the decay width converted to an axial-vector coupling $g$ for
each decay using the Goldberger-Treiman relation. This information can be found in Refs.~[\refcite{dai,rfm98}] and will not
be repeated here.

\begin{table}[ht]
\tbl{\label{t:tab1} Experimental data on three $\Delta S = 0$ HSD. The units of $R$ are 10$^{-3}$ s$^{-1}$ for neutron decay
and $10^6 \textrm{s}^{-1}$ for the other decays.}
{\footnotesize
\begin{tabular}{
l
r@{.}l@{$\pm$}r@{.}l r@{.}l@{$\pm$}r@{.}l r@{.}l@{$\pm$}r@{.}l
} \hline
&
\multicolumn{4}{c}{$n p$} &
\multicolumn{4}{c}{$\Sigma^+ \Lambda$} &
\multicolumn{4}{c}{$\Sigma^- \Lambda$} \\
\hline
$R$ &
1 & 1291 & 0 & 0010 & 0 & 249 & 0 & 062 & 0 & 387  & 0 & 018 \\
$\alpha_{e\nu}$ &
$-$0 & 0766 & 0 & 0036 & $-$0 & 35 & 0 & 15 & $-$0 & 404 & 0 & 044 \\
$\alpha_e$ &
$-$0 & 08559 & 0 & 00086 & \multicolumn{4}{c}{} & \multicolumn{4}{c}{} \\
$\alpha_\nu$ &
0 & 983 & 0 & 004 & \multicolumn{4}{c}{} & \multicolumn{4}{c}{} \\
$A$ &
\multicolumn{4}{c}{} & \multicolumn{4}{c}{} & 0 & 07 & 0 & 07 \\
$B$ &
\multicolumn{4}{c}{} & \multicolumn{4}{c}{} & 0 & 85 & 0 & 07 \\
$g_1/f_1$ &
1 & 2695 & 0 & 0029 & \multicolumn{4}{c}{} & \multicolumn{4}{c}{} \\ \hline
\end{tabular}
}
\end{table}

\begin{table}[ht]
\tbl{\label{t:tab2} Experimental data on five $|\Delta S| = 1$ HSD. The units of $R$ are $10^6 \textrm{s}^{-1}$.}
{\footnotesize
\begin{tabular}{
l
r@{.}l@{$\pm$}r@{.}l r@{.}l@{$\pm$}r@{.}l
r@{.}l@{$\pm$}r@{.}l r@{.}l@{$\pm$}r@{.}l
r@{.}l@{$\pm$}r@{.}l
} \hline
&
\multicolumn{4}{c}{$\Lambda p$} &
\multicolumn{4}{c}{$\Sigma^- n$} &
\multicolumn{4}{c}{$\Xi^- \Lambda$} &
\multicolumn{4}{c}{$\Xi^- \Sigma^0$} &
\multicolumn{4}{c}{$\Xi^0 \Sigma^+$} \\
\hline
$R$ &
3 & 161 & 0 & 058 & 6 & 88 & 0 & 24 & 3 & 44 & 0 & 19 &
0 & 53 & 0 & 10 & 0 & 93 & 0 & 14 \\
$\alpha_{e\nu}$ &
$-$0 & 019 & 0 & 013 & 0 & 347 & 0 & 024 & 0 & 53 & 0 & 10 &
\multicolumn{4}{c}{} & \multicolumn{4}{c}{} \\
$\alpha_e$ &
0 & 125 & 0 & 066 & $-$0 & 519 & 0 & 104 & \multicolumn{4}{c}{} &
\multicolumn{4}{c}{} & \multicolumn{4}{c}{} \\
$\alpha_\nu$ &
0 & 821 & 0 & 060 & $-$0 & 230 & 0 & 061 & \multicolumn{4}{c}{} &
\multicolumn{4}{c}{} & \multicolumn{4}{c}{} \\
$\alpha_B$ &
$-$0 & 508 & 0 & 065 & 0 & 509 & 0 & 102 & \multicolumn{4}{c}{} &
\multicolumn{4}{c}{} & \multicolumn{4}{c}{} \\
$A$ &
\multicolumn{4}{c}{} & \multicolumn{4}{c}{} & 0 & 62 & 0 & 10 &
\multicolumn{4}{c}{} & \multicolumn{4}{c}{} \\
$g_1/f_1$ &
0 & 718 & 0 & 015 & $-$0 & 340 &
0 & 017 & 0 & 25 & 0 & 05 &
1 & 287 & 0 & 158 & 1 & 32 & 0 & 22 \\ \hline
\end{tabular}
}
\end{table}

\section{Fits to experimental data}

At this point we can proceed to perform detailed comparisons with the experimental data of Tables \ref{t:tab1} and
\ref{t:tab2} through a number of fits. The experimental data which are used are the decay rates and the spin and angular
correlation coefficients of the HSD listed in these tables. The value of the ratio $g_1/f_1$ is not used since it is not an
independent measurement. For the processes $\Xi^-\to \Sigma^0 e^- \overline \nu_e$ and $\Xi^0\to \Sigma^+ e^- \overline
\nu_e$, however, we do use $g_1/f_1$ because no information on the angular coefficients is available yet. The theoretical
expressions for the integrated observables can be found in Refs.~[\refcite{rfm04,acosta}]. In order to have a reliably
determination of $V_{ud}$ and $V_{us}$, we systematically incorporate radiative corrections into the various integrated
observables and include the momentum-transfer contributions of the form factors. As for the decuplet baryons, we use the 
axial coupling $g$.

The parameters to be fitted are those arising out of the $1/N_c$ expansions of the baryon operators, namely, $v_{1-3}$ for
$f_1$ and $a$, $b$, $d$, $c_{1-4}$ for $g_1$. We use the values of $f_2$ and $g_2$ in the limit of exact SU(3) flavor symmetry.
The matrix elements $V_{ud}$ and $V_{us}$ are allowed to be free parameters too. Hereafter, the quoted errors of the best fit
parameters will be from the $\chi^2$ fit only, and will not include any theoretical uncertainties.

\subsection{$\Delta S=0$ fit and the extraction of $V_{ud}$}

As stated above, we can proceed to make an investigation of the $\Delta S=0$ processes only, namely, $n \rightarrow p$ and
$\Sigma^\pm \rightarrow \Lambda$ semileptonic decays and the strong decays $\Delta \to N \pi$, $\Xi^*\to\Lambda \pi$,
$\Sigma^*\to \Sigma \pi$, and $\Xi^* \to \Xi \pi$. We proceed right away to evaluate first-order SB effects in $g_1$, which
depends on the parameters $a$, $b$, $d$, $c_3$ and $c_4$, as introduced in Eq.~(\ref{eq:ds0f}). The fit produces
$a=0.89\pm 0.02$, $b=-0.21\pm 0.04$, $d=-0.06\pm 0.01$, $c_3=-0.08\pm 0.01$, $c_4=0.01 \pm 0.01$, and
\begin{equation}
V_{ud}= 0.9741 \pm 0.0005, \label{eq:vudf}
\end{equation}
with $\chi^2=8.62$ for 5 degrees of freedom. The results are consistent with expectations. The leading parameter $a$ is of
order unity, $b\sim 1/N_c$ is small compared to $a$, $d\sim 1/N_c^2$ is rather small and $c_3,c_4 \sim \epsilon/N_c$ are
consistent with first-order breaking $\epsilon$ divided by $N_c$.

\subsection{$|\Delta S|=1$ fit and the extraction of $V_{us}$}

The $|\Delta S|=1$ sector of HSD has been studied in detail in Ref.~[\refcite{rfm04}]. The experimental
information used is listed in Table \ref{t:tab2}. An interesting case of study was the incorporation not only of first-order
breaking effects into $g_1$, but also second-order breaking corrections into $f_1$, according to the Ademollo-Gatto theorem.
Therefore, the parameters which enter into play are $v_{1-3}$ of $f_1$, Eq.~(\ref{eq:vds1}), simultaneously with $a^\prime$,
$b^\prime$, and $c_{3-4}$ of $g_1$, Eq.~(\ref{eq:ds1f}), while $f_2$ and $g_2$ remain fixed by exact SU(3) symmetry. The best
fit parameters are\cite{rfm04} $v_1=0.03\pm 0.04$, $v_2=0.03\pm 03$, $v_3=0.01\pm 0.01$, $a^\prime=0.72\pm 0.03$,
$b^\prime=-0.08\pm 0.01$, $c_3=0.03\pm 0.02$, $c_4=0.05 \pm 0.02$, and
\begin{equation}
V_{us}= 0.2199 \pm 0.0026, \label{eq:vusf}
\end{equation}
with $\chi^2=17.85$ for 10 degrees of freedom. Again, the best fit parameters are as expected from the $1/N_c$ expansion
predictions. All the physical quantities of interest corresponding to this fit (observables, form factors, pattern of
symmetry breaking) can be found in this reference and will not be repeated here. Let us discuss our findings in the
concluding section.

\section{Discussion}

In our analysis, although we have performed a series of fits under several assumptions in the context of exact and broken
SU(3) symmetry, we summarize only salient results. Currently, the Particle Data Group\cite{part} recommends the values
$V_{ud}=0.9738 \pm 0.005$, which comes from superallowed $0^+\to 0^+$ beta decays and neutron beta decay, and
$V_{us}=0.2200 \pm 0.0026$, which comes mainly from $K_{e3}$ decays. From HSD we are able to determine
$V_{ud}=0.9741 \pm 0.005$ and $V_{us}=0.2199 \pm 0.0026$, which are comparable to the former. When the values obtained here
are combined with $V_{ub}$ of Ref.~[\refcite{part}], one finds $|V_{ud}|^2 + |V_{us}|^2 + |V_{ub}|^2 =0.9972 \pm 0.0016$,
which fails to satisfy unitarity by 1.8 sigma.

While our fit may not turn out to be definite, there is nevertheless a clear trend. We confirm that deviations from the exact
SU(3) limit, accounted for in the form factors $f_1$ and $g_1$, are indeed important in order to reliably determine the CKM
matrix elements $V_{ud}$ and $V_{us}$ from HSD. In particular, our determination of $V_{us}$ agrees well with recent
determinations from hadronic $\tau$ decays\cite{gamiz} and $K_{e3}$ decays\cite{lai}. However, recent works on $K_{e3}$
decays\cite{ktev,sher,batt} and HSD\cite{cabb} find higher values of $V_{us}$, which suggest a better agreement with
unitarity. This fact is not a drawback of our analysis and only states the need of more work, both theoretical and
experimental. The $1/N_c$ expansion may provide useful guidance in this task.

\section*{Acknowledgments}
R.F.M.\ is grateful to the organizers for making this workshop a successful one and to ECT, Trento, for the warm hospitality
extended to him. This work was partially supported by Consejo Nacional de Ciencia y Tecnolog{\'\i}a and Fondo de Apoyo a la
Investigaci\'on (Universidad Aut\'onoma de San Luis Potos{\'\i}), Mexico.

\end{document}